\documentclass{article}

\usepackage{arxiv}

\usepackage[utf8]{inputenc} 
\usepackage[T1]{fontenc}    
\usepackage{hyperref}       
\usepackage{url}            
\usepackage{booktabs}       
\usepackage{amsfonts}       
\usepackage{nicefrac}       
\usepackage{microtype}      
\usepackage[ruled,vlined]{algorithm2e}
\usepackage{todonotes}
\usepackage{graphicx,caption,subcaption}

\usepackage{comment}

\usepackage{amsmath}
\usepackage{amssymb}
\usepackage{natbib}
\usepackage{xcolor}

\newtheorem{theorem}{Theorem}
\newtheorem{definition}{Definition}
\newtheorem{lemma}{Lemma}

\newcommand{\Expect}[1]{\mathbb{E} \left[{#1}\right]}
\newcommand{\Expects}[2]{\mathbb{E}_{{#1}} \left[{#2}\right]}

\newcommand{\bx}{\mathbf{x}}
\newcommand{\bX}{\mathbf X}

\newcommand{\by}{\mathbf y}
\newcommand{\btheta}{\boldsymbol\theta}

\newcommand{\bxi}{\boldsymbol\xi}

\newcommand{\st}{\btheta} 

\title{Efficient and Generalizable Tuning Strategies for Stochastic Gradient MCMC}

\author{%
  Jeremie Coullon \\
  Department of Mathematics and Statistics\\
  Lancaster University\\
  United Kingdom \\
  \texttt{j.coullon@lancaster.ac.uk} \\
   \And
   Leah South \\
   School of Mathematical Sciences \\
   Queensland University of Technology \\
   Australia \\
   \texttt{l1.south@qut.edu.au} \\
   \AND
   Christopher Nemeth \\
  Department of Mathematics and Statistics\\
  Lancaster University\\
  United Kingdom \\
  \texttt{c.nemeth@lancaster.ac.uk} \\
}

\begin{document}

\maketitle

\begin{abstract}
Stochastic gradient Markov chain Monte Carlo (SGMCMC) is a popular class of algorithms for scalable Bayesian inference. However, these algorithms include hyperparameters such as step size or batch size that influence the accuracy of estimators based on the obtained posterior samples. As a result, these hyperparameters must be tuned by the practitioner and currently no principled and automated way to tune them exists. Standard MCMC tuning methods based on acceptance rates cannot be used for SGMCMC, thus requiring alternative tools and diagnostics. We propose a novel bandit-based algorithm that tunes the SGMCMC hyperparameters by minimizing the Stein discrepancy between the true posterior and its Monte Carlo approximation. We provide theoretical results supporting this approach and assess various Stein-based discrepancies. We support our results with experiments on both simulated and real datasets, and find that this method is practical for a wide range of applications.
\end{abstract}











\section{INTRODUCTION}

Most MCMC algorithms contain user-controlled hyperparameters which need to be carefully selected to ensure that the MCMC algorithm explores the posterior distribution efficiently. Optimal tuning rates for many popular MCMC algorithms such the random-walk \citep{gelman1997weak} or Metropolis-adjusted Langevin algorithms \citep{roberts1998optimal} rely on setting the tuning parameters according to the Metropolis-Hastings acceptance rate. Using metrics such as the acceptance rate, hyperparameters can be optimized on-the-fly within the MCMC algorithm using adaptive MCMC \citep{andrieu2008tutorial,vihola2012robust}. However, in the context of stochastic gradient MCMC (SGMCMC), there is no acceptance rate to tune against and the trade-off between bias and variance for a fixed computation budget means that tuning approaches designed for target invariant MCMC algorithms are not applicable.


\textbf{Related work} Previous adaptive SGMCMC algorithms have focused on embedding ideas from the optimization literature within the SGMCMC framework, e.g. gradient preconditioning \citep{li2016preconditioned}, RMSprop \citep{chen2016bridging} and Adam \citep{kim2020stochastic}. However, all of these algorithms still rely on hyperparameters such as learning rates and subsample sizes which need to be optimized. To the best of our knowledge, no principled approach has been developed to optimize the SGMCMC hyperparameters. In practice, users often use a trial-and-error approach and run multiple short chains with different hyperparameter configurations and select the hyperparameter setting which minimizes a metric of choice, such as the kernel Stein discrepancy \citep{nemeth2020stochastic} or cross-validation \citep{izmailov2021bayesian}. However, this laborious approach is inefficient and not guaranteed to produce the best hyperparameter configuration.  

\textbf{Contribution} In this paper we propose a principled adaptive SGMCMC scheme that allows users to tune the hyperparameters, e.g. step-sizes $h$ (also known as the learning rate) and data subsample size $n$. Our approach provides an automated trade-off between bias and variance in the posterior approximation for a given computational time budget. Our adaptive scheme uses a multi-armed bandit algorithm to select SGMCMC hyperparameters which minimize the Stein discrepancy between the approximate and true posterior distributions. The approach only requires a user-defined computational budget as well as the gradients of the log-posterior, which are already available to us via the stochastic gradient MCMC algorithm. A second contribution in this paper is a rigorous assessment of existing tuning methods for SGMCMC, which to our knowledge is not present in the literature. 


\section{BACKGROUND}

\subsection{Stochastic Gradient Langevin Algorithm}

We are interested in sampling from a target density $\pi(\st)$, where for some parameters of interest $\st \in \mathbb{R}^d$ the unnormalized density is of the form $\pi(\st) \propto \exp\{ -U(\st)\}$. We assume that the potential function $U(\st)$ is continuous and differentiable almost everywhere. If we have independent data, $y_1,\ldots,y_N$ then $\pi(\st)\propto p(\st)\prod_{i=1}^N f(y_i|\st)$ is the posterior density, where $p(\st)$ is the prior density and $f(y_i|\st)$ is the likelihood for the $i$th observation. In this setting, we can define $U(\st)=\sum_{i=1}^N U_i(\st)$, where $U_i(\st)= - \log f(y_i|\st)-(1/N)\log p(\st)$.  

We can sample from $\pi(\st)$ by simulating a stochastic process that has $\pi$ as its stationary distribution. Under mild regularity conditions, the Langevin diffusion \citep{Roberts:1996,pillai2012optimal} has $\pi$ as its stationary distribution, however, in practice it is not possible to simulate the Langevin diffusion exactly in continuous time and instead we sample from a discretized version. That is, for a small time-interval $h>0$, the Langevin diffusion has approximate dynamics given by
\begin{equation} \label{eq:sgld}
    \st_{k+1} = \st_{k} - \frac{h}{2} \nabla U(\st(t)) + \sqrt{h} \bxi_k, \quad k=0,\ldots,K
\end{equation}
where $\bxi_k$ is a vector of $d$ independent standard Gaussian random variables. In the large data setting, we replace $\nabla U(\st)$ with an unbiased estimate $\nabla \tilde{U}(\st) = \frac{N}{n} \sum_{i \in \mathcal{S}_n} \nabla U_i(\st)$, calculated using a subsample of the data of size $n<<N$, where $\mathcal{S}_n$ is a random sample, without replacement, from $\{1,\ldots,N\}$. This algorithm is known as the stochastic gradient Langevin dynamics \citep[SGLD,][]{Welling:2011}.

In this paper we present our adaptive stochastic gradient MCMC scheme in the context of the SGLD algorithm for simplicity of exposition. However, our proposed approach is readily generalizable to all other stochastic gradient MCMC methods, e.g. stochastic gradient Hamiltonian Monte Carlo \citep{chen2014stochastic}. Details of the general class of stochastic gradient MCMC methods presented under the \textit{complete recipe} framework are given in \citet{ma2015complete}. See Section \ref{sec:appendix_sgmcmc_samplers} of the Supplementary Material for a summary of the SGMCMC algorithms used in this paper.

\subsection{Stein Discrepancy}


We define $\tilde{\pi}$ as the empirical distribution generated by the stochastic gradient MCMC algorithm \eqref{eq:sgld}. We can define a measure of how well this distribution approximates our target distribution of interest, $\pi$, by defining a discrepancy metric between the two distributions. Following \cite{gorham2015measuring} we consider the Stein discrepancy 
\begin{align}
    D(\tilde{\pi},\pi) = \sup_{\phi \in \mathcal{F}} \left| \mathbb{E}_{\tilde{\pi}}[\underbrace{-\nabla_{\st}U(\st)^{\top} \phi(\st)   + \nabla_{\st}^{\top} \phi(\st)}_{\text{Stein operator: }\mathcal{A}_{\pi}\phi(\st)}]\right|
\end{align}
where $\phi: \mathbb{R}^d \rightarrow \mathbb{R}^d$ is any smooth function in the Stein set $\mathcal{F}$ which satisfies Stein's identity $\Expects{\pi}{\mathcal{A}_{\pi}\phi(\st)}=0$ for all $\phi \in \mathcal{F}$.

\textbf{Kernel Stein Discrepancy} To obtain an analytic form of the Stein discrepancy, \cite{liu2016kernelized,chwialkowski2016kernel} introduced the kernelized Stein discrepancy (KSD) which 
has a closed form solution
\begin{align}
\label{eq:ksd}
    \mathrm{KSD}(\tilde{\pi},\pi) 
= \sqrt{\Expects{\tilde{\pi}(\st)\tilde{\pi}(\st')}{k_{\pi}(\st,\st')}}
\end{align}

where 
\begin{align*}
 k_{\pi}(\st,\st^\prime) = & \nabla_{\st} U(\st)^{\top} \nabla_{\st'} U(\st') k(\st,\st') - \nabla_{\st} U(\st)^{\top} \nabla_{\st'} k(\st,\st') \\
-  & \nabla_{\st'} U(\st')^{\top} \nabla_{\st} k(\st,\st') + \nabla_{\st}^{\top} \nabla_{\st'} k(\st,\st^\prime). 
\end{align*}
Interested readers can find technical details on the corresponding choice of $\mathcal{F}$ in \cite{gorham2017measuring}. The kernel $k$ must be positive definite, which is a condition satisfied by most popular kernels, including the Gaussian and Mat\'ern kernels. \cite{gorham2017measuring} recommend using the inverse multi-quadratic kernel, $k(\st,\st^\prime) = (c^2 + ||\st-\st^\prime||_2^2)^\beta$, which they prove detects non-convergence when $c>0$ and $\beta \in (-1,0)$.  

\textbf{Finite Set Stein Discrepancy} KSD is a natural discrepancy measure for stochastic gradient MCMC algorithms as $\pi(\btheta)$ is only required up to a normalization constant and the gradients of the log-posterior density are readily available. The drawback of KSD is that the computational cost is quadratic in the number of samples. Linear versions of the KSD \citep{liu2016kernelized} are an order of magnitude faster, but the computational advantage is outweighed by a significant decrease in the accuracy of the Stein estimator. 

\cite{jitkrittum2017linear} propose a linear-time Stein discrepancy, the Finite Set Stein Discrepancy (FSSD), which utilizes the \textit{Stein witness function} $g(\st^\prime):=\mathbb{E}_{\st\sim\tilde{\pi}}[-\nabla_{\st}U(\st)^{\top} k(\st,\st^\prime) + \nabla_{\st}^{\top} k(\st,\st^\prime)]$. The function $g$ can be thought of as witnessing the differences between $\tilde{\pi}$ and $\pi$, where a discrepancy in the region around $\st$ is indicated by large $|g(\st)|$. The Stein discrepancy is essentially then measured via the flatness of $g$, where the measure of flatness can be computed in linear time.  
The key to FSSD is to use real analytic kernels $k$, e.g, Gaussian kernel, which results in $g_1,\ldots,g_d$ also having a real analytic form. If $g_i \neq 0$ then this implies almost surely that $g_i(\mathbf{v}_1),\ldots,g_i(\mathbf{v}_J)$ are not zero for a finite set of test locations $V=\{\mathbf{v}_1,\ldots,\mathbf{v}_J\}$. Under the same assumptions as KSD, FSSD is defined as,
\begin{align}
\label{eq:fssd}
\mathrm{FSSD}(\tilde{\pi},\pi):=\sqrt{\frac{1}{dJ}\sum_{i=1}^d\sum_{j=1}^{J}g_i^2(\mathbf{v}_j)}.
\end{align}

Theorem 1 of \cite{jitkrittum2017linear} guarantees that $\mathrm{FSSD}^2=0$ if and only if $\tilde{\pi}=\pi$ for any choice of test locations $\{\mathbf{v}\}_{j=1}^J$. However, some test locations will result in an improved  test power for finite samples and so, following \cite{jitkrittum2017linear}, we optimize the test locations by first sampling them from a Gaussian fit to the posterior samples and then use gradient ascent so that they maximise the FSSD.

\section{HYPERPARAMETER LEARNING}

In this section we introduce an automated and generally-applicable approach to learning the user-controlled parameters of a stochastic gradient MCMC algorithm, which throughout we will refer to as hyperparameters. For example, in the case of SGLD, this would be the stepsize parameter $h$ and batch size $n$, or in the case of stochastic gradient Hamiltonian Monte Carlo, this would also include the number of leap frog steps. Our adaptive scheme relies on multi-armed bandits \citep{slivkins2019introduction} to identify the optimal setting for the hyperparameters such that, for a given time budget, the 
selected parameters minimize the Stein discrepancy, and therefore maximize the accuracy of the posterior approximation. Our proposed approach, the Multi-Armed MCMC Bandit Algorithm (MAMBA), works by sequentially identifying and pruning, i.e. removing, poor hyperparameter configurations in a principled, automatic and online setting to speed-up hyperparameter learning. The MAMBA algorithm can be used within any stochastic gradient MCMC algorithm and only requires the user to specify the training budget and the number of hyperparameter sets. 



\subsection{Multi-Armed Bandits with Successive Halving}





Multi-armed bandits are a class of algorithms for sequential decision-making that iteratively select actions from a set of possible decisions. These algorithms can be split into two categories: 1) \textit{best arm identification} in which the goal is to identify the action with the highest average reward, and 2) \textit{exploration vs. exploitation}, where the goal is to maximize the cumulative reward over time \citep{bubeck2012regret}. In the best-arm
identification setting, an action, \textit{aka arm}, is selected and  produces a reward, where the reward is drawn from a fixed probability distribution corresponding to the chosen arm. At the end of the exploration phase, a single arm is chosen which maximizes the expected reward. This differs from the typical multi-armed bandit setting where the strategy for selecting arms is based on minimizing cumulative regret \citep{lattimore2020bandit}. 

 The \textit{successive halving} algorithm \citep{karnin2013almost,jamieson2016non} is a multi-armed bandit algorithm based on best arm identification. Successive halving learns the best hyperparameter settings, i.e. the best arm, using a principled early-stopping criterion to identify the best arm within a set level of confidence, or for a fixed computational budget. In this paper, we consider the fixed computational budget setting, where the algorithm proceeds as follows: 1) uniformly allocate a computational budget to a set of arms, 2) evaluate the performance of all arms against a chosen metric, 3) promote the best $1/\eta$ of arms to the next stage, where typically $\eta=2 \ \text{or} \ 3$, and prune the remaining arms from the set. The process is repeated until only one arm remains. As the total computational budget is fixed, pruning the least promising arms allows the algorithm to allocate exponentially more computational resource to the most promising hyperparameter sets.

   
   
    
    
    


\subsection{Tuning Stochastic Gradients with a Multi-Armed MCMC Bandit Algorithm (MAMBA)}\label{ssec:MAMBA}

We describe our proposed algorithm, MAMBA, to tune the hyperparameters of a generic stochastic gradient MCMC algorithm. For ease of exposition, we present MAMBA in the context of the SGLD algorithm \eqref{eq:sgld}, where a user tunes the step size $h$ and batch size $n$. Details on other SGMCMC algorithms can be found in Appendix \ref{sec:appendix_sgmcmc_samplers}. We present MAMBA as the following three stage process:

\textbf{Initialize} In our multi-armed bandit setting we assume $M$ possible stochastic gradient MCMC  hyperparameter configurations, which we refer to as \textit{arms}. Each arm $s$ in the initial set $S_0 = \{1,\ldots,M\}$
represents a hyperparameter tuple $\phi_s=(h_s,n_s)$. The hyperparameters in the initial set are chosen from a uniform grid. 
 
 \textbf{Evaluate and Prune} At each iteration of MAMBA, $i = 0, 1, \ldots$, each arm $s$ is selected from the set $S_i$ and the $s^\text{th}$ SGLD algorithm is run for $r_i$ seconds using the hyperparameter configuration $\phi_s$. Each arm is associated with a reward $\nu_s$ that measures the quality of the posterior approximation. We use the negative Stein discrepancy as the reward function that we aim to maximize. Specifically, we calculate the Stein discrepancy from the SGMCMC output using KSD \eqref{eq:ksd} or FSSD \eqref{eq:fssd}, i.e. $\nu_s = -\mathrm{KSD}(\tilde{\pi}_{s},\pi)$ or $\nu_s = -\mathrm{FSSD}(\tilde{\pi}_{s},\pi)$. Without loss of generality, we can order the set of arms $S_i$ by their rewards, i.e.  $\nu_1 \geq \nu_2 \geq \ldots \geq\nu_{M}$, where $\nu_1$ is the arm with the optimal reward at each iteration of MAMBA. The top $100/\eta \%$ of arms in $S_i$ with the highest rewards are retained to produce the set $S_{i+1}$. The remaining arms are pruned from the set and not evaluated again at future iterations. 

\textbf{Reallocate time} Computation time allocated to the pruned samplers is reallocated to the remaining samplers, $r_{i+1} = \eta r_i$. As a result, by iteration $i$, each of the remaining SGLD samplers has run for a time budget of $R = r_0 + \eta r_0 + \eta^2r_0 + ... + \eta^{i-1}r_0$ seconds, where $r_0$ is the time budget for the first MAMBA iteration. This process is repeated for a total of $ \lfloor\log_{\eta}M\rfloor$ MAMBA iterations. We use a $\log_\eta$ base as we are dividing the number of arms by $\eta$ at every iteration. Furthermore, we use a floor function for the cases where the initial number of arms $M$ is not a power of $\eta$. The MAMBA algorithm is summarized in Algorithm \ref{alg:example}.

\begin{algorithm}[H]
   \caption{MAMBA}
   \label{alg:example}
   \textbf{Input:} Initial number of configurations $M$, total time budget $T$ and pruning rate $\eta$ (default $\eta=3$).  Sample $M$ hyperparameter configurations and store in the set $S_0$. 
   
   \For{$i=0 \hspace{2mm} \mathrm{to} \hspace{2mm} \lfloor \log_{\eta}M\rfloor - 1$}{
    $r_i = \frac{T}{|S_i|\lfloor \log_{\eta}M \rfloor}$
    
    Run each SGLD sampler using \eqref{eq:sgld} for time budget of $r_i$ seconds. \\
    Calculate KSD or FSSD for each sampler using  \eqref{eq:ksd} or \eqref{eq:fssd}, respectively. \\
    Let $S_{i+1}$ be the set of $\lfloor |S_i| / \eta \rfloor$ samplers with lowest KSD/FSSD.
    }
\end{algorithm}

\textbf{Guarantees} It is possible that MAMBA will eliminate the optimal hyperparameter set during one of the arm-pruning phases. Through examination of the $1-1/2\eta$ quantile, we show in Theorem \ref{thm} that we can bound the probability that MAMBA, using negative KSD as the reward function, will incorrectly prune the best hyperparameter configuration and provide a bound on the maximum computation budget required for MAMBA to identify the optimal hyperparameter setting.

\begin{definition}
Let $s \in \{2,\ldots,M\}$ be an arm with reward $\nu_s$, then we define the suboptimality gap between $\nu_s$ and the optimal reward $\nu_1$ as $\alpha_s := \nu_1 - \nu_s$, and we define $H_2 := \max_{s \neq 1} s/\alpha_s^2$ as the complexity measure, see \cite{audibert2010best} for details. 
\end{definition}

\begin{theorem}
\label{thm}
i) MAMBA correctly identifies the best hyperparameter  configuration for a stochastic gradient MCMC algorithm with probability at least
\[1 - 
(2\eta-1) \log_\eta M \cdot\exp{\left(-\frac{\eta T}{4\sigma^2_{\mathrm{KSD}} H_2 (\log_\eta M+1)}\right)},
\]
where $\sigma^2_{\mathrm{KSD}} = \max_{s \in S}\mathrm{Var}_{\tilde{\pi}_{s}}(\Expects{\tilde{\pi}_{s}}{k_\pi(\btheta,\btheta')})$.

ii) For a probability of at least $1-\delta$ that MAMBA will successively identify the optimal hyperparameter set, MAMBA requires a computational budget of 
\[
T = O\left(\sigma^2_{\mathrm{KSD}} \log_\eta M\log\left(\frac{(2\eta-1)\log_\eta M}{\delta}\right)\right).
\]
\end{theorem}

A proof of Theorem \ref{thm} is given in Appendix \ref{app:Proof} and builds on the existing work of \cite{karnin2013almost} for fixed-time best-arm identification bandits. Theorem \ref{thm} highlights the contribution of KSD variance in identifying the optimal arm. In particular, the total computation budget depends on the arm with the largest KSD variance. 

\subsection{Practical Guidance for Using MAMBA}\label{sec:practical_guidance_SATGrad}

\textbf{Time budgets} In the case of sampling algorithms like SGMCMC, the relevant budgets are total computation time or number of iterations. If we view sampling algorithms as a trade-off between statistical accuracy and computation time, then the goal is to identify the hyperparameters that produce the best Monte Carlo estimates under a given time constraint. Using a time budget, rather than number of iterations, means we can choose the best data subsample size in a principled way, as a smaller subsample will lead to a faster algorithm, but increase the Monte Carlo error. Additionally, using a time budget to select the optimal hyperparameters ties the statistical efficiency to the available computational power and implementation, e.g., programming language, hardware, etc.

\textbf{Estimating KSD/FSSD} Calculating KSD/FSSD using \eqref{eq:ksd} or \eqref{eq:fssd} requires the gradients of the log-posterior and the SGMCMC samples. Typically, one would calculate the KSD/FSSD using fullbatch gradients (i.e. using the entire dataset) on the full chain of samples. However, as we only use SGMCMC when the dataset is large, this would be a computationally expensive approach. Two natural solutions are to i) use stochastic gradients \citep{gorham2020stochastic}, calculated through subsampling, or ii) use a thinned chain of samples. We investigate both options in terms of KSD/FSSD accuracy in Appendix \ref{sec:appendix_SH_tuning_details}. We find that using the stochastic KSD/FSSD produces results similar to the fullbatch KSD/FSSD. However, calculating the KSD/FSSD for a large number of high dimensional samples is computationally expensive, so for our experimental study in Section \ref{sec:experimental_study} we use fullbatch gradients with thinned samples. This leads to lower variance KSD/FSSD estimates at a reasonable computational cost. Note that fullbatch gradients are only used for MAMBA iterations and not SGMCMC iterations. We find that this does not significantly increase the overall computational cost as for each iteration of MAMBA there are thousands of SGMCMC iterations.

\textbf{Alternative metrics} Stein-based discrepancies are a natural metric to assess the quality of the posterior approximation as they only require the SGMCMC samples and log-posterior gradients. However, alternative metrics to tune SGMCMC can readily be applied within the MAMBA framework. For example, there is currently significant interest in understanding uncertainty in neural networks via metrics such as expected calibration error (ECE), maximum calibration error (MCE) (\cite{Guo:2017}), and out-of-distribution (OOD) tests \citep{Lakshminarayanan:2017}. These metrics have the advantage that they are more scalable to very high dimensional problems, compared to the KSD \citep{Gong:sliced_KSD_2020}. 
As a result, although KSD is a sensible choice when aiming for posterior accuracy, alternative metrics may be more appropriate for some problems, for example, in the case of very high-dimensional deep neural networks.

\section{EXPERIMENTAL STUDY}
\label{sec:experimental_study}

In this section we illustrate MAMBA on three different models. We optimize the hyperparameters using MAMBA (Alg. \ref{alg:example}) and compare against a grid search approach and a heuristic method. The initial arms in MAMBA are set as an equally spaced grid over batch sizes and step sizes (and number of leapfrog steps for SGHMC). The heuristic method fixes the step size to be inversely proportional to the dataset size, i.e. \(h=\frac{1}{N}\) \citep{Brosse:2018}. For both the grid search and heuristic approaches, we use a \(10\%\) batch size throughout. Note that MAMBA is the only method able to estimate both step size and batch size. Full details of the experiments can be found in Appendix \ref{sec:appendix_details_experiments}. Code to replicate the experiments can be found at \href{https://github.com/RedactedForReview}{https://github.com/RedactedForReview}. All experiments were carried out on a laptop CPU (MacBook Pro 1.4 GHz Quad-Core Intel Core i5). For each example, the figures show results over a short number of tuning iterations and tables give results for longer runs.


\subsection{Logistic Regression}\label{sec:LR}

We consider logistic regression on a simulated dataset with 10 dimensions and 1 million data points (details of the model and prior are in Appendix \ref{sec:appendix_log_reg_model}). We sample from the posterior using six samplers: SGLD, SGLD with control variates \citep[SGLD-CV,][]{Baker:2017}, and stochastic gradient Hamiltonian Monte Carlo \citep[SGHMC,][]{chen2014stochastic}, SGHMC-CV, stochastic gradient Nose Hoover Thermostats (SGNHT), and SGNHT-CV \citep{ding2014bayesian}.

For MAMBA, we set $R=1sec$ (i.e.: the running time of the longest sampler). We point out that this time budget is small compared to what would be used by most practitioners. However, this example illustrates the MAMBA methodology and compares it against a full MCMC algorithm which provides us with "ground-truth" posterior samples. To calculate the KSD/FSSD efficiently, we thin the samples and use fullbatch gradients.

\begin{figure*}[h]
\centering
\includegraphics[width=0.9\linewidth]{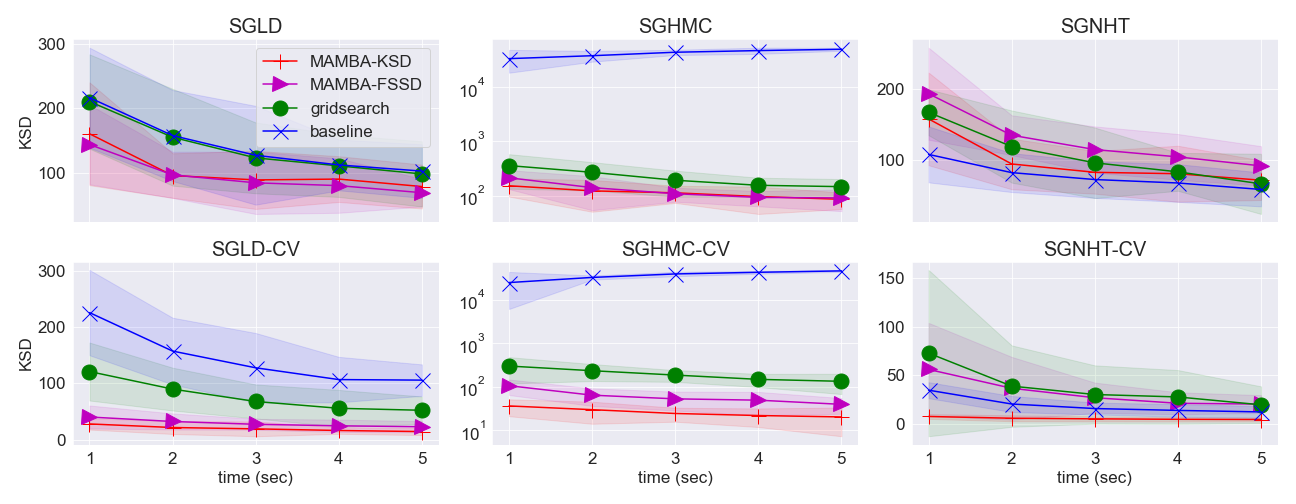}
\caption{KSD curves for the six samplers applied to a logistic regression model.}
\label{fig:LR_ksd_curves}
\end{figure*}

In Figure \ref{fig:LR_ksd_curves}, we plot the KSD calculated from the posterior samples for each of the tuning methods. We calculated the KSD curves for ten independent runs and plotted the mean curve along with a confidence interval (two standard deviations). The optimal hyperparameters given by each method can be found in Table \ref{tab:LR_hyperparameters} of Appendix \ref{sec:appendix_log_reg_model}. Our results from Figure \ref{fig:LR_ksd_curves} show that optimizing the hyperparameters with MAMBA, using either KSD or FSSD, produces samples that have the lowest KSD over all but one of the six samplers. For the SGNHT sampler, the heuristic approach gives the lowest KSD, however, as shown in Table \ref{tab:LR_hyperparameters} in Appendix \ref{sec:appendix_log_reg_model}, MAMBA-FSSD finds an optimal step size of $h=N^{-1}$, which coincides with step size given by the heuristic approach. Therefore, the difference in KSD from these two methods is a result of the batch size, which when taking into account computation time, MAMBA-FSSD finds $1\%$ to be optimal, whereas the heuristic method does not learn the batch size and this is fixed at $10\%$. Ignoring computation time, a larger batch size is expected to produce a better posterior approximation. However, it is interesting to note that for the five out of six samplers where MAMBA performs the best in terms of KSD, MAMBA chooses an optimal batch size of $1\%$.

For this simulated data example with only 1 million samples we can compare the posterior accuracy of the SGMCMC algorithms against the \textit{ground-truth} using NumPyro's (\cite{phan2019composable}, \cite{bingham2018pyro}) implementation of NUTS (\cite{hoffman2014no}) on the full dataset for 20K iterations (after a burn-in of 1K iterations). We then calculate the relative error in the posterior standard deviation for each sampler: $\xi(\hat{\sigma}):= ||\hat{\sigma}-\sigma_{\text{NUTS}}||_2 / || \sigma_{\text{NUTS}}||_2$. The results are given in Table \ref{tab:LR_std_estimates} and further results including predictive accuracy on a test dataset and the number of samples obtained within the time budget are given in Table \ref{tab:LR_std_estimates_full} of Appendix \ref{sec:appendix_log_reg_model}. We find that the MAMBA-optimized samplers perform the best in terms of KSD. As a result, the Monte Carlo estimates of the posterior standard deviations generally perform well. We tested each sampler by running each sampler for 10 seconds.


\begin{table*}
\caption{\textbf{Logistic regression}. For each tuning method and each SGMCMC sampler we report the relative standard deviation error and the KSD. We abbreviate MAMBA-KSD and MAMBA-FSSD to M-KSD and M-FSSD, respectively.}
\label{tab:LR_std_estimates}
\centering
\begin{tabular}{ p{2.0cm} p{2.0cm} p{1.6cm} p{1.6cm} p{1.6cm} p{1.6cm} p{1.6cm} p{1.6cm}}
  & & \textbf{SGLD} & \textbf{SGLD-CV} & \textbf{SGHMC} & \textbf{SGHMC-CV} & \textbf{SGNHT} & \textbf{SGNHT-CV}\\
\hline \\
 
 \textbf{M-KSD} 
  & KSD & 66 & \textbf{13} & 85 & \textbf{18} & 69 & \textbf{3} \\
  & $\xi(\hat{\sigma}) \times 10^{2}$ & 28.3 & \textbf{5.2} & 107.7 & \textbf{8.4} & 55.7 & \textbf{0.8}  \\
   \midrule

 \textbf{M-FSSD} 
  & KSD & \textbf{58} & 23 & \textbf{56} & 43 & 68 & 11 \\
  & $\xi(\hat{\sigma}) \times 10^{2}$ & 68.5 & 5.9 & 82.5 & 26.3 & 102.5 & 2.0 \\
   \midrule

 \textbf{Grid} 
  & KSD & 106 & 38 & 174 & 131 & 73 & 12 \\
  & $\xi(\hat{\sigma}) \times 10^{2}$ & \textbf{12.0} & 12.4 & \textbf{34.5} & 31.2 & \textbf{15.0} & 10.5 \\
   \midrule

 \textbf{Heuristic} 
  & KSD & 100 & 102 & 53,972 & 51,565 & \textbf{51} & 9 \\
  & $\xi(\hat{\sigma}) \times 10^{2}$ & 12.1 & 27.5 & 3,000.3 & 3,084.0 & 71.4 & 20.4  \\
   
\end{tabular}
\end{table*}

\subsection{Probabilistic Matrix Factorization}\label{sec:PMF}

We consider the probabilistic matrix factorization model \citep{salakhutdinov2008bayesian} on the MovieLens dataset\footnote{available at \href{https://grouplens.org/datasets/movielens/100k/}{https://grouplens.org/datasets/movielens/100k/}} \citep{movielens_2016}, which contains $100$K ratings for $1,682$ movies from $943$ users (see Appendix \ref{sec:appendix_PMF_model} for model details). We optimize the hyperparameters for six samplers: SGLD, SGLD-CV, SGHMC, SGHMC-CV, SGNHT, and SGNHT-CV.  We use a similar setup as for logistic regression and tune each sampler using MAMBA with KSD, grid search, and the heuristic method. Details are given in Appendix \ref{sec:appendix_PMF_details}.


\begin{figure*}[h]
\centering
\includegraphics[width=0.9\linewidth]{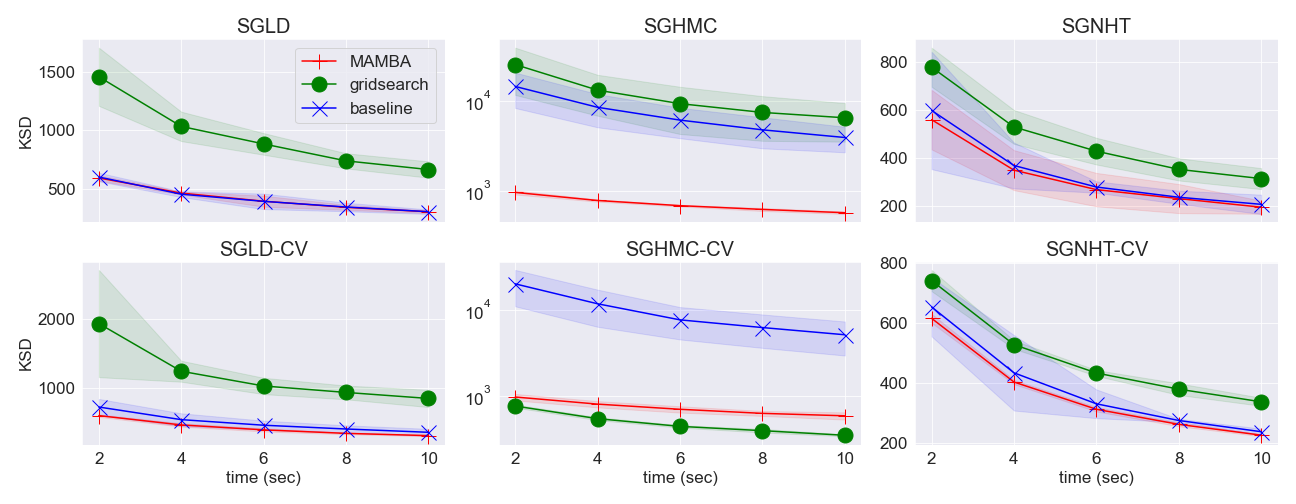}
\caption{KSD curves for the six samplers applied to the probabilistic matrix factorization model.}
\label{fig:PMF_ksd_curves}
\end{figure*}

From Figure \ref{fig:PMF_ksd_curves} we can see that the samplers tuned using MAMBA tend to outperform the ones tuned with the other two methods. We also test the quality of the posterior samples against NumPyro's (\cite{phan2019composable}, \cite{bingham2018pyro}) implementation of NUTS (\cite{hoffman2014no}), which produces 10K samples with 1K samples as burn-in. This state of the art sampler obtains high quality samples but is significantly more computationally expensive, taking around six hours on a laptop CPU. We estimate the posterior standard deviations using these samples and treat them as the ground-truth.  We run each SGMCMC sampler for 20 seconds, and estimate the standard deviation after removing the burn-in. We estimate the posterior standard deviation for each sampler and show the relative errors and KSD in Table \ref{tab:PMF_uncertainty_tests} (further results are given in Table \ref{tab:PMF_uncertainty_tests_full} in Appendix \ref{sec:appendix_PMF_details}). We find that MAMBA consistently identifies hyperparameters that give the lowest KSD, but that for some samplers the heuristic approach gives a lower error on the estimated standard deviation. This could be due to the random realisation of the SGMCMC chain; however, while accuracy in standard deviation is fast to compute, as a metric it is not as useful as KSD, which measures the quality of the full distribution and not just the accuracy of the second moment.




\begin{table*}
\caption{\textbf{Probabilistic Matrix Factorization}. For each tuning method and each sampler we report KSD and the relative error of the standard deviation estimates.}
\label{tab:PMF_uncertainty_tests}
\centering

\begin{tabular}{ p{2.0cm} p{2.0cm} p{1.6cm} p{1.6cm} p{1.6cm} p{1.6cm} p{1.6cm} p{1.6cm}}
  & & \textbf{SGLD} & \textbf{SGLD-CV} & \textbf{SGHMC} & \textbf{SGHMC-CV} & \textbf{SGNHT} & \textbf{SGNHT-CV}\\
\hline \\
 
 \textbf{MAMBA} 
  & KSD & \textbf{213} & \textbf{231} & \textbf{438} & \textbf{543} & \textbf{163} & \textbf{205}  \\
  & $\xi(\hat{\sigma}) \times 10^{2}$ & \textbf{69.2} & \textbf{72.1} & 79.9  & 83.7 & 40.8 & \textbf{38.5}  \\
   \midrule
   
 \textbf{Grid} 
  & KSD & 429 & 546 & 3,180 & 4,289 & 170 & 221  \\
  & $\xi(\hat{\sigma}) \times 10^{2}$ & 119.2 & 133.0 & 51.2 & 55.6 & 44.2 & 46.8   \\
   \midrule
 
 \textbf{Heuristic} 
  & KSD & 237 & 284 & 3,942 & 4,546 & 164 & 210 \\
  & $\xi(\hat{\sigma}) \times 10^{2}$ & 71.7 & 75.0 & \textbf{50.3}  & \textbf{53.2} & \textbf{40.6} & 38.7 \\
  
\end{tabular}
\end{table*}

\subsection{Bayesian Neural Network}\label{sec:BNN}

In this section we consider a feedforward Bayesian neural network with two hidden layers on the MNIST dataset \citep{Lecun2010} (see Appendix \ref{sec:appendix_NN_model} for details). Here we tune six samplers: SGLD, SGLD-CV, SGHMC, SGHMC-CV, SGNHT, and SGNHT-CV.

\begin{figure*}[h]
\centering
\includegraphics[width=0.9\linewidth]{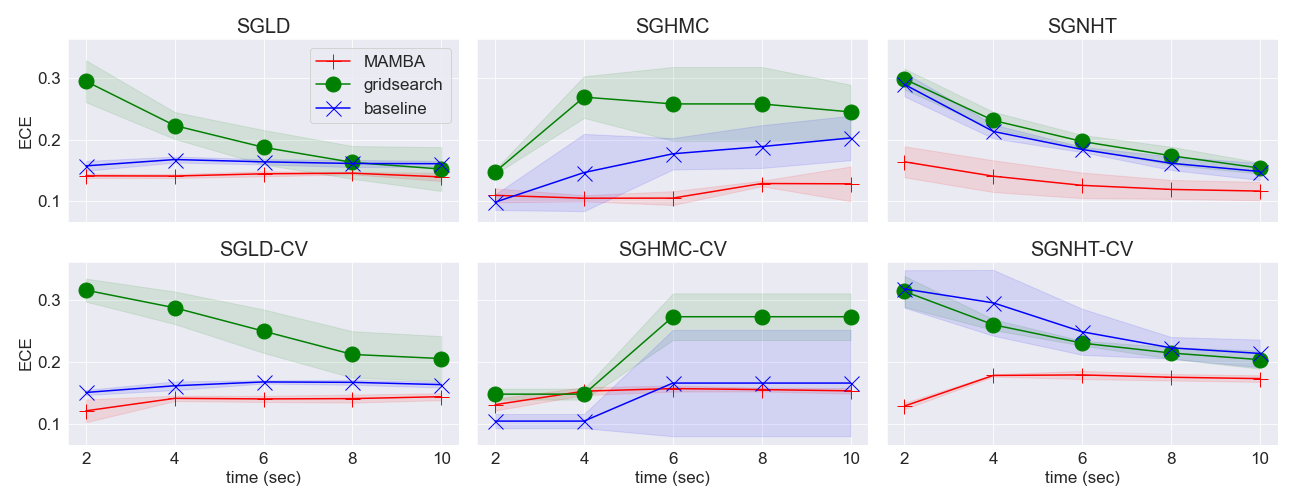} 
\caption{ECE curves for the six samplers applied to the Bayesian neural network model.}
\label{fig:NN_ksd_curves}
\end{figure*}

For this example, as with the previous examples, we tune MAMBA using KSD, however, we validate the accuracy of the various tuning approaches against expected calibration error (ECE) and maximum calibration error (MCE) plotted in Figure \ref{fig:NN_ksd_curves}. We find that the samplers tuned using MAMBA tend to outperform the other approaches in terms of ECE. We assess the performance of the MAMBA-optimized samplers over a longer time budget and run the samplers for 300 seconds starting from the maximum aposteriori value. We then remove the visible burn-in and calculate the ECE and MCE to compare the quality of the posterior samples. We report the results in Table \ref{tab:NN_uncertainty_tests}, where ECE and MCE are reported as percentages (lower is better).

\begin{table*}
\caption{\textbf{Bayesian Neural Network}. For each tuning method and each sampler we report the ECE and MCE (as percentages).}
\label{tab:NN_uncertainty_tests}
\centering

\begin{tabular}{ p{2.0cm} p{2.0cm} p{1.6cm} p{1.6cm} p{1.6cm} p{1.6cm} p{1.6cm} p{1.6cm}}
  & & \textbf{SGLD} & \textbf{SGLD-CV} & \textbf{SGHMC} & \textbf{SGHMC-CV} & \textbf{SGNHT} & \textbf{SGNHT-CV}\\
\hline \\
 
 \textbf{MAMBA} 
  & ECE ($\%$) & 1.0 & 0.9 & \textbf{0.7} & \textbf{0.7} & 9.3 & \textbf{0.9}   \\
  & MCE ($\%$) & 36.4 & \textbf{15.7} & \textbf{47.1} & \textbf{21.3} & 45.7 & \textbf{27.4}  \\
   \midrule
   
 \textbf{Grid} 
  & ECE ($\%$) & 14.6 & 8.8 & 20.1 & 25.1 & \textbf{5.4} & 7.7  \\
  & MCE ($\%$) & 42.1 & 40.7 & 65.5 & 55.2 & \textbf{42.2} & 42.3  \\
   \midrule
 
 \textbf{Heuristic} 
  & ECE ($\%$) & \textbf{0.8} & \textbf{0.7} & 50.9 & 40.8 & 6.2 & 7.0   \\
  & MCE ($\%$) & \textbf{23.3} & 22.0 & 71.6 & 74.8 & 43.2 & 51.5 \\
  
\end{tabular}
\end{table*}



Overall, the results in Table \ref{tab:NN_uncertainty_tests} show that MAMBA-optimized samplers tend to perform best in terms of KSD and when not the best they produce results which are very close to the best performing method. For all samplers, MAMBA finds an optimal batch size of $1\%$, which is ten times smaller than the batch size of the other methods and therefore results in a faster and highly accurate algorithm. For SGNHT both MAMBA and grid search found a step size that was slightly too large ($\log_{10}(h)=-4.5$ and $\log_{10}(h)=-4$ respectively) which caused the sampler to lose stability for longer chains. In contrast, the sampler tuned using the heuristic method is the only one that remained stable. As a result we re-ran these two tuning methods for a grid with smaller step sizes: \(\{-5. , -5.5, -6. , -6.5, -7. , -7.5\}\). This smaller grid allowed the two tuning algorithms to find a stable step size ($\log_{10}(h)=-5$ for both methods), and so this slight decrease in step size was enough to make the sampler stable. We note that there exists samplers with more stable numerical methods such as the BADODAB sampler which solves the same diffusion as SGNHT but with a more stable splitting method \citep{Leimkuhler_2016_badodab}. Such samplers might be easier to tune with MAMBA or grid search.




\section{DISCUSSION AND FUTURE WORK}\label{sec:Discussion}
\textbf{Final remarks} In this paper we have proposed a multi-armed bandit approach to estimate the hyperparameters for any stochastic gradient MCMC algorithm. Our approach optimizes the hyperparameters to produce posterior samples which accurately approximate the posterior distribution within a fixed time budget. We use Stein-based discrepancies as natural metrics to assess the quality of the posterior approximation.

The generality of the MAMBA algorithm means that alternative metrics, such as predictive accuracy, can easily be employed within MAMBA as an alternative to a Stein-based metric. Whilst not explored in this paper, it is possible to apply MAMBA beyond the stochastic gradient MCMC setting and directly apply MAMBA to standard MCMC algorithms, such as Hamiltonian Monte Carlo, to estimate the MCMC hyperparameters.  

Finally, in this paper we performed a systematic study of different SGMCMC tuning methods for various models and samplers, which to our knowledge is the first rigorous comparison of these methods. While these alternative approaches can work well they are only able to tune the step size parameter, and unlike MAMBA, they do not tune the batch size or other SGMCMC hyperparameters, such as the number of leap frog steps. 



\textbf{Limitations and future work} A limitation of this method is that computing the KSD can be expensive when there are many posterior samples. One solution we explored in this paper is to use the FSSD as a linear-time metric. In the case of KSD, we significantly lowered the cost of this by thinning the Markov chain, but the KSD remains an expensive metric to compute. The KSD also suffers from the curse of dimensionality \citep{Gong:sliced_KSD_2020}, though our results show that the KSD gave good results even for our two high-dimensional problems. As a result, further work in this area should explore alternative discrepancy metrics which are both scalable in sample size and dimension. For example, scalable alternatives to KSD, such as sliced KSD \citep{Gong:sliced_KSD_2020}, would be more appropriate for very high-dimensional problems.

\bibliographystyle{bibstyle}
\bibliography{main}

\clearpage
\appendix

\begin{center}
    \Large{\textbf{Appendix}}
\end{center}

\section{PROOF OF THEOREM \ref{thm}}\label{app:Proof}

\begin{lemma}
\label{ksd-lemma}
Assume $\{\btheta_i\}_{i=1}^P$ are P samples from $\tilde{\pi}$ and $k(\btheta, \btheta')$ is a positive definite kernel in the Stein class of $\tilde{\pi}$ and $\pi$ . If $\Expects{\tilde{\pi}}{k_\pi(\btheta,\btheta')^2} < \infty$, and $\tilde{\pi} \neq \pi$, then for a Monte Carlo approximation of the kernel Stein discrepancy \eqref{eq:ksd}, we have
$$
\sqrt{P}\left(\widehat{\mathrm{KSD}^2}(\tilde{\pi},\pi)-\mathrm{KSD}^2(\tilde{\pi},\pi)\right) \xrightarrow[]{\text{D}} \mathcal{N}(0,\sigma^2),
$$
where $\sigma^2 = \mathrm{Var}_{\tilde{\pi}}(\Expects{\tilde{\pi}}{k_\pi(\btheta,\btheta')})$.
\end{lemma}

This lemma follows directly from Sections 5.5.1 and 5.5.2 of \cite{serfling2009approximation}.

\begin{lemma}
\label{hoeffding-lemma}
Let $s$ be an arm from the set of arms $S_i=\{1,2,\ldots\}$ at iteration $i$ of the MAMBA algorithm. We let $s=1$ be the optimal arm with expected reward $\bar{\nu}_1$ and we assume that the optimal arms was not eliminated at iteration $i-1$ of the MAMBA algorithm. We then have for any arm $s \in S_i$ with estimated reward $\hat{\nu}_s$, 
\[
P(\bar{\nu}_1 < \hat{\nu}_s) \leq \exp\left(-\frac{\alpha_s^2 r_i}{2 \sigma_s^2}\right).
\]
\end{lemma}
\textit{Proof:} Using the CLT result from Lemma \ref{ksd-lemma} we assume that $\hat{\nu}_s$ is an unbiased estimate of the reward for arm $s$ with sub-Gaussian proxy $\sigma_s^2$, then by the Hoeffding inequality we have
\begin{align*}
    P\left(\hat{\nu}_s-\bar{\nu}_s \geq \alpha_s\right) \leq \exp\left(-\frac{\alpha_s^2 r_i}{2 \sigma_s^2}\right),
\end{align*}
where $\alpha_s := \bar{\nu}_1 - \bar{\nu}_s$ and therefore $P\left(\hat{\nu}_s-\bar{\nu}_s \geq \alpha_s\right) = P(\bar{\nu}_1 < \hat{\nu}_s)$ and the lemma follows. $\hfill \square$

\begin{lemma}
\label{arm-elimination}
The probability that the best arm is eliminated at iteration $i$ of MAMBA (Algorithm \ref{alg:example}) is at most
\[
(2\eta-1) \exp{\left(-\frac{\eta T}{4\sigma_s^2(\log_\eta M+1)}\cdot\frac{\alpha_{s_i}^2}{s_i}\right)},
\]
where $s_i=M/\eta^{i+1}$.
\end{lemma}
\textit{Proof:} This result follows a similar process to Lemma 4.3 from \cite{karnin2013almost} but for a general $\eta$. If the best arm is removed at iteration $i$, then there must be at least $1/\eta$ arms in $S_i$ (i.e. $\frac{1}{\eta}|S_i|=M/\eta^{i+1}$) with empirical reward larger than that of the best arm (i.e. a KSD score lower than the arm with the best possible KSD score). If we let $S_i'$ be the set of arms in $S_i$, excluding the $|S_i|/2\eta=M/2\eta^{i+1}$ arms with largest reward, then the empirical reward for at least $|S_i'|/(2\eta-1)=M/2\eta^{i+1}$ arms in $|S_i'|$ must be greater than the best arm at iteration $i$.

Let $N_i$ be the number of arms in $S_i'$ with empirical reward greater than the reward of the optimal arm, then by Lemma \ref{hoeffding-lemma} we have,
\begin{align*}
    \Expect{N_i} &= \sum_{s \in S_i'} P(\bar{\nu}_1<\hat{\nu}_s) \leq \sum_{s \in S_i'} \exp{\left(-\frac{\alpha_s^2 r_i}{2\sigma_s^2}\right)} \leq \sum_{s \in S_i'} \exp{\left(-\frac{\alpha_s^2}{2\sigma_s^2}\cdot\frac{T}{|S_i|(\log_\eta (M)+1)}\right)} \\
    & \leq |S_i'|\max_{s \in S_i'}\exp{\left(-\frac{\alpha_s^2}{2\sigma^2}\cdot\frac{\eta^i T}{M(\log_\eta M+1)}\right)} \leq |S_i'|\exp{\left(-\frac{\eta T}{4\sigma^2(\log_\eta M+1)}\cdot\frac{\alpha_{s_i}^2}{s_i}\right)},
\end{align*}
where $\sigma^2=\max_{s\in S_i'}\sigma_s^2$ and the final inequality follows from the fact that there are at least $s_i-1$ arms that are not in $S_i'$ with reward greater than any arm in $S_i'$. Applying Markov's inequality we can obtain,

\[
P(N_i>|S_i'|/(2\eta-1)) \leq (2\eta-1)\Expect{N_i}/|S_i'| \leq (2\eta-1) \exp{\left(-\frac{\eta T}{4\sigma^2(\log_\eta M+1)}\cdot\frac{\alpha_{s_i}^2}{s_i}\right)}.
\] 
$\hfill \square$

Using Lemmas \ref{hoeffding-lemma} and \ref{arm-elimination} we can now prove Theorem \ref{ksd-lemma}.

\textit{Proof:} The algorithm cannot exceed to the budget of $T$ (in our case $T$ is given in seconds). If the best arm survives then the algorithm succeeds as all other arms must be eliminated after $\log_\eta M$ iterations. Finally, using Lemma \ref{arm-elimination} and a union bound, the best arm is eliminated in one of the $\log_\eta M$ iterations of the algorithm with probability at most
\begin{align*}
    (2\eta-1) & \sum_{i=1}^{\log_\eta (M)}\exp{\left(-\frac{\eta T}{4\sigma^2(\log_\eta (M)+1)}\cdot\frac{\alpha_{s_i}^2}{s_i}\right)} \\
    & \leq (2\eta-1) \log_\eta (M) \cdot\exp{\left(-\frac{\eta T}{4\sigma^2(\log_\eta (M)+1)}\cdot\frac{1}{\max_s s\alpha_s^{-2}}\right)} \\
     & \leq (2\eta-1) \log_\eta (M) \cdot\exp{\left(-\frac{\eta T}{4\sigma^2 H_2 (\log_\eta (M)+1)}\right)}. \\
\end{align*} 
This result completes the proof of Theorem \ref{thm}.
$  \hfill\square$

\section{TUNING METHODS}\label{sec:app_tuning}

\subsection{Grid search and heuristic method}

For grid search we run the sampler using the training data, and calculate the RMSE/log-loss/accuracy on the test dataset. To have a fair comparison to MAMBA (see Section \ref{sec:appendix_SH_tuning_details}), we always start the sampler from the maximum \textit{a posteriori} estimate (the MAP, found using optimization). As a result we need to add noise around this MAP or else the grid search tuning method will recommend the smallest step size available which results in the sampler not moving away from the starting point. This happens because the MAP has the smallest RMSE/ log-loss (or highest accuracy). To fix this we add Gaussian noise to the MAP, and report the scale of the noise for each model in Section \ref{sec:appendix_details_experiments}.

The heuristic method fixes the step size to be inversely proportional to the dataset size, i.e. \(h=\frac{1}{N}\) \citep{Brosse:2018}. For both the grid search and heuristic approaches, we use a \(10\%\) batch size throughout. 

\subsection{MAMBA}\label{sec:appendix_SH_tuning_details}

We investigate the tradeoffs involved in estimating the KSD from samples in MAMBA. We can estimate this using the stochastic gradients estimated in the SGMCMC algorithm. However we can also calculate the fullbatch gradients and use these to estimate the KSD. Although the latter option is too computationally expensive in the big data setting, we can also thin the samples to estimate the KSD which may result in the fullbatch gradients being computationally tractable.

\begin{figure*}[h!]
\centering
\includegraphics[width=1\linewidth]{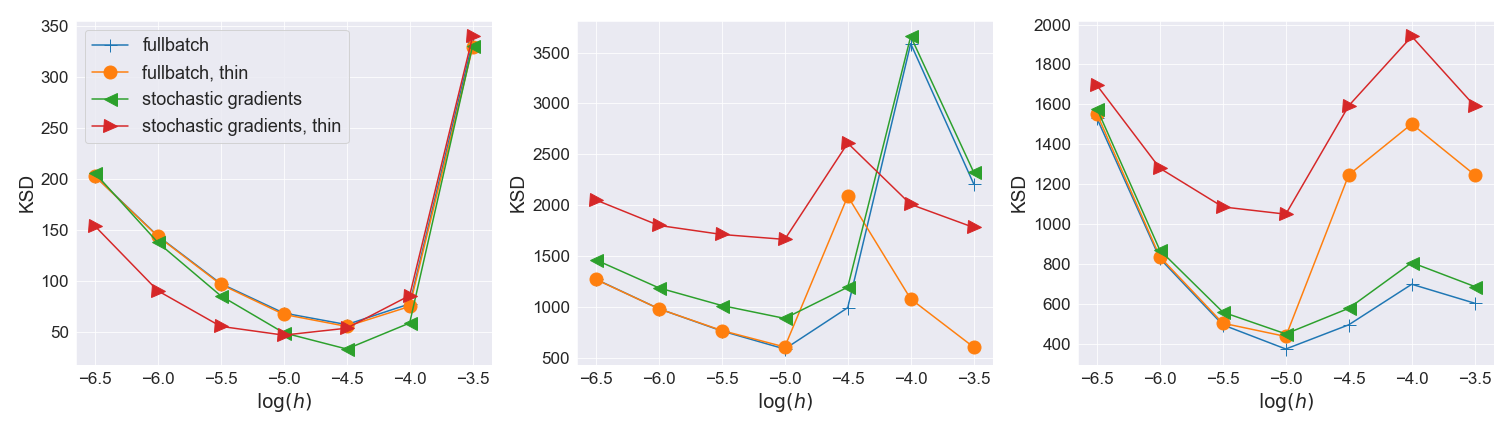}
\caption{Grid search for different step sizes using both fullbatch and stochastic-KSD for logistic regression, PMF, and NN (from left to right). The sampler used is SGLD}
\label{fig:KSD_test_step_size}
\end{figure*}

\begin{figure*}[h!]
\centering
\includegraphics[width=1\linewidth]{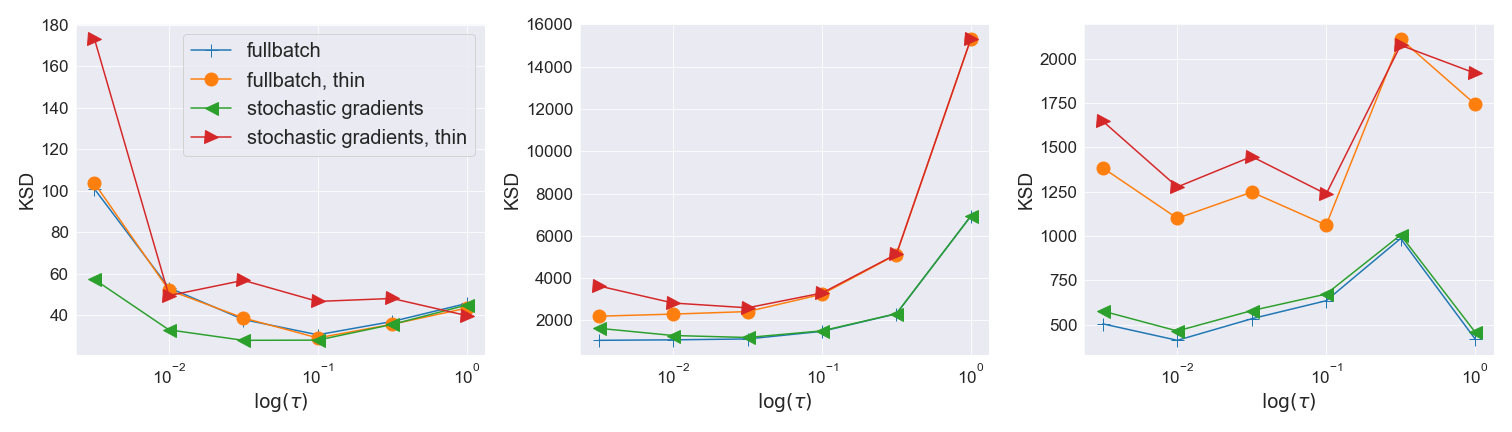}
\caption{Grid search for different batch sizes using both fullbatch and stochastic-KSD for logistic regression, PMF, and NN (from left to right). The sampler used is SGLD}
\label{fig:KSD_test_batch_size}
\end{figure*}

In Figure \ref{fig:KSD_test_step_size} we estimate the KSD of samples using the logistic regression model over a grid of step sizes. We run SGLD for the 3 models for 1 second and with a batch size of $1\%$. We estimate the KSD in 4 ways: fullbatch using all the samples, fullbatch using thinned samples (thin by 5), stochastic gradients using all samples, and stochastic gradients using thinned samples. In Figure \ref{fig:KSD_test_batch_size} we do the same but varying the batch size (and keeping the step size fixed to $h = 10^{-4.5}$. We can see that the KSD estimated using stochastic gradients and unthinned samples follows the fullbatch KSD well. However as calculating the KSD for many high dimensional samples is computationally expensive, we opt for using thinned fullbatch gradients in all our experiments.

\section{DETAILS OF EXPERIMENTS}\label{sec:appendix_details_experiments}

We use the SGMCMC samplers in the package \href{https://github.com/jeremiecoullon/SGMCMCJax}{SGMCMCJax} for the experiments, and we use NumPyro for the NUTS sampler (\cite{phan2019composable}, \cite{bingham2018pyro}).

\subsection{Logistic regression}\label{sec:appendix_log_reg_model}

\subsubsection{Model}

Consider a binary regression model where $\by=\{y_i\}_{i=1}^N$ is a vector of  $N$ binary responses and $\bX$ is a $N \times d$ matrix of covariates. If $\st$ is a $d-$dimensional vector of model parameters, then the likelihood function for the logistic regression model is,
$$
p(\by, \bX \ | \ \st) = \prod_{i=1}^N \left[\frac{1}{1+\exp(-\st^\top\bx_i)}\right]^{y_i}\left[1-\frac{1}{1+\exp(-\st^\top\bx_i)}\right]^{1-y_i}
$$
where $\bx_i$ is a $d-$dimensional vector for the $i$th observation. The prior distribution for $\st$ is a zero-mean Gaussian with covariance matrix $\boldsymbol{\Sigma}_{\st}=10\mathbf{I}_d$, where $\mathbf{I}_d$ is a $d \times d$ identity matrix.

\subsubsection{Grid search}
\label{sec:logistic_regression_grid_search}

For grid search we choose the step size using $14$ equally spaced step sizes (on a $log_{10}$ scale) that result in the best log-loss on a test dataset: \(\{-1. , -1.5, -2. , -2.5, -3. , -3.5, -4. , -4.5, -5. , -5.5, -6. , -6.5, -7. , -7.5\}\). To tune SGHMC we use the same step size grid with two leapfrog values: 5 and 10. We fix the batch size ratio to be \(10\%\) for the grid search tuning method as well as the baseline.

We start from the MAP with Gaussian noise (scale: $\sigma=0.2$) and run the samplers for $5,000$ for each grid point.


\subsubsection{MAMBA}\label{sec:appendix_log_reg_model_SH}

We use the same grid for step sizes as grid search and use a grid of four batch sizes: \(100\%\), \(10\%\), \(1\%\), and \(0.1\%\). To calculate the KSD and FSSD we thin the samples by 10 and calculate the fullbatch gradients. For FSSD-opt we samples 10 test locations from a Gaussian fit to the samples and optimize them using Adam.

\subsubsection{Results}

Table \ref{tab:LR_hyperparameters} gives the hyperparameters used to produce the runs in Table \ref{tab:LR_std_estimates}.

\begin{table*}
\caption{Logistic regression: hyperparameters for the results in Table \ref{tab:LR_std_estimates}. The batch size is given by $\tau$: the percentage of the total number of data. Namely: batch size $n = \lfloor{\tau N/ 100}\rfloor$}
\label{tab:LR_hyperparameters}
\centering

\begin{tabular}{ p{2.2cm} p{2cm} p{1.6cm} p{1.6cm} p{1.6cm} p{1.6cm}}
  & & \textbf{MAMBA-KSD} & \textbf{MAMBA-FSSD} & \textbf{Grid Search} & \textbf{Heuristic}\\
\hline \\
 \textbf{SGLD} & $\log_{10}(h)$  & -6 & -5.5 & -6 & -6 \\
                 & $\tau$ ($\%$)  & 1  & 1  & 10 &  10 \\
 \midrule
 
 \textbf{SGLD-CV} & $\log_{10}(h)$  & -5 & -5 & -5 & -6 \\
     & $\tau$ ($\%$) & 0.1 & 1 & 10 &  10\\
                
 \midrule

  \textbf{SGHMC} & $\log_{10}(h)$  & -7 & -6 & -7 & -6 \\
     & $\tau$ ($\%$) & 0.1 & 1 & 10 &  10\\
     & $L$  & 10 & 5 & 10 & 10 \\
  
  \midrule
  \textbf{SGHMC-} & $\log_{10}(h)$  & -6.5 & -6 & -6.5 & -6 \\
   \textbf{CV}  & $\tau$ ($\%$) & 0.1 & 1 & 10 &  10\\
     & $L$  & 10 & 5 & 10 & 10 \\
  
 \midrule
  \textbf{SGNHT}  & $\log_{10}(h)$  & -7.5 & -6 & -7.5 & -6 \\
     & $\tau$ ($\%$) & 1 & 1 & 10 &  10\\
  
   \midrule
  \textbf{SGNHT-CV}  & $\log_{10}(h)$  & -5.5 & -5 & -4.5 & -6 \\
     & $\tau$ ($\%$) & 1 & 10 & 10 &  10\\
\end{tabular}

\end{table*}

\begin{table*}
\caption{Comparison of tuning methods for Logistic regression. For each tuning method and each sampler we report the relative error of the standard deviation estimates, the KSD, the predictive accuracy, and the number of samples. Note that the number of samples generated within a fixed time budget depends on the subsample size. We try two version of MAMBA, one with KSD as a metric, and the other with FSSD-opt}
\label{tab:LR_std_estimates_full}
\centering
\begin{tabular}{ p{2.2cm} p{2.5cm} p{1.6cm} p{1.6cm} p{1.6cm} p{1.6cm}}
  & & \textbf{MAMBA-KSD} & \textbf{MAMBA-FSSD} & \textbf{Grid Search} & \textbf{Heuristic}\\
\hline \\
 
 \textbf{SGLD} & $ \xi(\hat{\sigma}) \times 10^{2}$  & 28.3 & 68.5 & \textbf{12} & 12.1 \\
  & KSD & 66 & \textbf{58} & 106 & 100\\
 & Pred. acc. $(\%)$ &  93.9 & 93.9 & 93.9 & 93.9 \\
  & \# of samples & 22,255 & 21851 & 4005 & 3484 \\
 \midrule
 
 \textbf{SGLD-} & $ \xi(\hat{\sigma}) \times 10^{2}$  & \textbf{5.2} & 5.9 & 12.4 & 27.5 \\
  \textbf{CV} & KSD & \textbf{13} & 23 & 38 & 102 \\
  & Pred. acc. $(\%)$ & 93.9 & 93.9 & 93.9 & 93.9\\
  & \# of samples & 55,580 & 18,792 & 3,232 & 2,809  \\
 \midrule

  \textbf{SG-} & $ \xi(\hat{\sigma}) \times 10^{2}$ & 107.7 & 82.5 & \textbf{34.5} & 3,000.3 \\
   \textbf{HMC} & KSD & 85 & \textbf{56} & 174 & 53972 \\
  & Pred. acc. $(\%)$ &  93.9 & 93.9 & 93.9 & 92.7 \\
  & \# of samples & 28,268 & 5,145 & 435 & 428 \\
  
  \midrule
  \textbf{SGHMC-} & $ \xi(\hat{\sigma}) \times 10^{2}$ & \textbf{8.4} & 26.3 & 31.2 & 3,084.0 \\
   \textbf{CV} & KSD & \textbf{18} & 43 & 131 & 51,565 \\
  & Pred. acc. $(\%)$ & \textbf{93.9} & \textbf{93.9} & \textbf{93.9} & 92.7 \\
  & \# of samples & 24,001 & 4,194 & 355 & 346 \\

 \midrule
  \textbf{SGNHT} & $ \xi(\hat{\sigma}) \times 10^{2}$ & 55.7 & 102.5 & \textbf{15.0} & 71.4  \\
     & KSD & 69 & 68 & 73 & \textbf{51} \\
  & Pred. acc. $(\%)$ & 93.9 & 93.9 & 93.9 & 93.9 \\
  & \# of samples & 21,955 & 19,845 & 4,055  & 4,056  \\
  
   \midrule
  \textbf{SGNHT-} & $ \xi(\hat{\sigma}) \times 10^{2}$ & \textbf{0.8} & 2.0 & 10.5 & 20.4 \\
    \textbf{CV} & KSD & \textbf{3} & 11 & 12 & 9 \\
  & Pred. acc. $(\%)$ & 93.9 & 93.9 & 93.9 & 93.9 \\
  & \# of samples & 19,323 & 3,105 & 3,329 & 3,329 \\
\end{tabular}
\end{table*}

\subsection{Probabilistic Matrix Factorization}\label{sec:appendix_PMF_details}
\subsubsection{Model}\label{sec:appendix_PMF_model}

In this example, we will consider the MovieLens dataset \footnote{available at \href{https://grouplens.org/datasets/movielens/100k/}{https://grouplens.org/datasets/movielens/100k/}} \citep{movielens_2016} which contains $100,000$ ratings (taking values $\{1,2,3,4,5\}$) of $1,682$ movies by $943$ users, where each user has provided at least $20$ ratings. The data are already split into $5$ training and test sets ($80\%/20\%$ split) for a $5-$fold cross-validation experiment. 
Let $\mathbf{R} \in \mathbb{R}^{N \times M}$ be a matrix of observed ratings for $N$ users and $M$ movies where $R_{ij}$ is the rating user $i$ gave to movie $j$. We introduce matrices $\mathbf{U}$ and $\mathbf{V}$ for users and movies respectively, where $\mathbf{U}_i \in \mathbb{R}^d$ and $\mathbf{V}_j \in \mathbb{R}^d$ are $d-$dimensional latent feature vectors for user $i$ and movie $j$. The likelihood for the rating matrix is
$$
p(\mathbf{R}|\mathbf{U},\mathbf{V},\alpha) = \prod_{i=1}^N \prod_{j=1}^M \left[N(R_{ij}|\mathbf{U}_i^\top \mathbf{V}_j,\alpha^{-1})\right]^{I_{ij}}
$$
where $I_{ij}$ is an indicator variable which equals 1 if user $i$ gave a rating for movie $j$. The prior distributions for the users and movies are
$$
p(\mathbf{U}|\mathbf{\mu}_{\mathbf{U}},\Lambda_{\mathbf{U}}) = \prod_{i=1}^N N(\mathbf{U}_i|\mathbf{\mu}_{\mathbf{U}},\Lambda_{\mathbf{U}}^{-1}) \ \ \ \ \mbox{and} \ \ \ \ p(\mathbf{V}|\mathbf{\mu}_{\mathbf{V}},\Lambda_{\mathbf{V}}) = \prod_{j=1}^M N(\mathbf{V}_j|\mathbf{\mu}_{\mathbf{V}},\Lambda_{\mathbf{V}}^{-1}),
$$
with prior distributions on the hyperparameters (where $\mathbf{W}= \mathbf{U}$ or $\mathbf{V}$) given by,
$$
\mathbf{\mu}_{\mathbf{W}} \sim N(\mathbf{\mu}_{\mathbf{W}}|\mathbf{\mu}_0,\Lambda_{\mathbf{W}}) \ \ \ \ \mbox{and} \ \ \ \ \Lambda_{\mathbf{W}} \sim \mbox{Gamma}(a_0,b_0).
$$
The parameters of interest in our model are then $\st=(\mathbf{U},\mathbf{\mu}_{\mathbf{U}},\Lambda_{\mathbf{U}},\mathbf{V},\mathbf{\mu}_{\mathbf{V}},\Lambda_{\mathbf{V}})$ and the hyperparameters for the experiments are $\boldsymbol{\tau}=(\alpha,\mu_0,a_0,b_0)=(3,0,4,5)$. We are free to choose the size of the latent dimension and for these experiments we set $d=20$.

\subsubsection{Grid search}

To run grid search we use a grid of 12 step sizes in grid search (on a $log_{10}$ scale): \(\{-2. , -2.5, -3. , -3.5, -4. , -4.5, -5. , -5.5, -6. , -6.5, -7. , -7.5\}\). For SGHMC we also try two values of leapfrog steps: 5 and 10. 

We start from the MAP with Gaussian noise (scale: $\sigma=1$) and run $2,000$ iteration per grid point.

\subsubsection{MAMBA}

We use a time budget of $R=10sec$ (time of longest running sampler), and the same step size grid as for gridsearch: \(\{-2. , -2.5, -3. , -3.5, -4. , -4.5, -5. , -5.5, -6. , -6.5, -7. , -7.5 \}\). We also use a grid for batch sizes: \(100\%, 10\%, 1\%\). For SGHMC we try two values of leapfrog steps: 5 and 10.

\subsubsection{Results}

We show in Table \ref{tab:PMF_hyperparameters} the hyperparameters for the runs in Table \ref{tab:PMF_uncertainty_tests}.

\begin{table*}
\caption{PMF: Hyperparameters for the results in Table \ref{tab:PMF_uncertainty_tests}. The batch size is given by $\tau$: the percentage of the total number of data. Namely: batch size $n = \lfloor{\tau N / 100}\rfloor$}
\label{tab:PMF_hyperparameters}
\centering
\begin{tabular}{ p{2.5cm} p{1.5cm} p{2cm} p{2cm} p{2cm}}
  & & \textbf{MAMBA-KSD} &  \textbf{Grid Search} & \textbf{Heuristic}\\
\hline \\
 \textbf{SGLD} & $\log_{10}(h)$  & -5 & -3.5 & -4.9 \\
                 & $\tau$ ($\%$)  & 1 & 10 &  10 \\
 \midrule
 
 \textbf{SGLD-CV} & $\log_{10}(h)$    & -5 & -3.5 & -4.9 \\
     & $\tau$ ($\%$) & 1   & 10 &  10\\
                
 \midrule

  \textbf{SGHMC} & $\log_{10}(h)$   & -6 & -5 &  -4.9\\
     & $\tau$ ($\%$) & 1 & 10 &  10\\
     & $L$  & 5 & 10 & 10 \\
  \midrule
  \textbf{SGHMC-CV} & $\log_{10}(h)$  & -6 & -5 &  -4.9\\
     & $\tau$ ($\%$) &  1  & 10 &  10\\
     & $L$  & 5 & 10 & 10 \\
  
 \midrule
  \textbf{SGNHT}  & $\log_{10}(h)$  & -5 & -5.5 & -4.9 \\
     & $\tau$ ($\%$) & 10  & 10 &  10\\
  
   \midrule
  \textbf{SGNHT-CV}  & $\log_{10}(h)$  & -5 & -5.5 & -4.9 \\
     & $\tau$ ($\%$) & 10 & 10 &  10\\
\end{tabular}
\end{table*}

\begin{table*}
\caption{Comparison of tuning methods for PMF. For each tuning method and each sampler we report the relative error of the standard deviation estimates, the RMSE on test dataset, and the number of samples.}
\label{tab:PMF_uncertainty_tests_full}
\centering
\begin{tabular}{ p{1.7cm} p{2.5cm} p{3cm} p{3cm} p{3cm} }
  & & \textbf{MAMBA} & \textbf{Grid Search} & \textbf{Heuristic} \\
\hline \\
 
 \textbf{SGLD} & $ \xi(\hat{\sigma}) \times 10^{2}$  & \textbf{69.2} & 119.2 & 71.7\\
 & KSD &  \textbf{213} & 429 & 237  \\
 & RMSE & \textbf{1.13} & 1.25 & 1.13 \\
 & \# of samples & 15,681 & 10,099 & 9,946\\
 \midrule
 
 \textbf{SGLD-CV} & $ \xi(\hat{\sigma}) \times 10^{2}$  & \textbf{72.1} & 133.0 & 75.0\\
 & KSD & \textbf{231} & 546 & 284  \\
 & RMSE & \textbf{1.13} & 1.25 & \textbf{1.13} \\
 & \# of samples & 11,774 & 6,827 & 6,897 \\
 \midrule

 \textbf{SGHMC} & $ \xi(\hat{\sigma}) \times 10^{2}$ &  79.9 & 51.2 & \textbf{50.3}  \\
 & KSD &  \textbf{438} & 3,180 & 3,942 \\
 & RMSE & \textbf{1.13} & 1.25 & 1.25 \\
 & \# of samples & 3,503 & 1,184 & 1,021  \\
 
 \midrule
 \textbf{SGHMC-CV}& $ \xi(\hat{\sigma}) \times 10^{2}$ & 83.7 & 55.6 & \textbf{53.2}  \\
 & KSD &  \textbf{543} & 4,289 & 4,546  \\
 & RMSE & \textbf{1.10} & 1.25 & 1.25 \\
 & \# of samples &  2,045 & 848 & 847  \\
 
 \midrule
  \textbf{SGNHT} & $ \xi(\hat{\sigma}) \times 10^{2}$ & 40.8 & 44.2 & \textbf{40.6} \\
 & KSD &  \textbf{163} & 170 & 164 \\
 & RMSE & 1.12 &  1.15 & \textbf{1.11} \\
 & \# of samples &  9687 & 9743 & 9683 \\
 
 \midrule 
   \textbf{SGNHT-CV} & $ \xi(\hat{\sigma}) \times 10^{2}$ &  \textbf{38.5}  & 46.8 & 38.7 \\
 & KSD & \textbf{205} & 221 & 210 \\
 & RMSE & 1.12 & 1.16 & \textbf{1.11} \\
 & \# of samples &  6,527 & 5,930 & 6,546 \\
 
\end{tabular}
\end{table*}

\subsection{Neural Network}\label{sec:appendix_NN_details}

\subsubsection{Model}\label{sec:appendix_NN_model}

We consider the problem of multi-class classification on the popular MNIST dataset\footnote{\href{https://creativecommons.org/licenses/by-sa/3.0/}{CC A-SA 3.0}} \citep{Lecun2010}. The MNIST dataset consists of a collection of images of handwritten digits from zero to nine, where each image is represented as $28 \times 28$ pixels. We model the data using a two layer Bayesian neural network with 100 hidden variables (using the same setup as \cite{chen2014stochastic}). We fit the neural network to a training dataset containing $60,000$ images and the goal is to classify new images as belonging to one of the ten categories. The test set contains $10,000$ handwritten images, with corresponding labels.
Let $y_i$ be the image label taking values $y_i \in \{0,1,2,3,4,5,6,7,8,9\}$ and $\mathbf x_i$ is the vector of pixels which has been flattened from a $28 \times 28$ image to a one-dimensional vector of length 784. If there are $N$ training images, then $\mathbf X$ is a $N \times 784$ matrix representing the full dataset of pixels. We model the data as categorical variables with the probability mass function,
\begin{equation}
    p(y_i= k \, | \, \st, \mathbf x_i) = \beta_k(\st,\mathbf x_i),
    \label{eq:beta}
\end{equation}
where $\beta_k(\st, \mathbf x_i)$ is the $k$th element of $\beta(\st, \mathbf x_i) = \sigma \left( \sigma \left( \mathbf x_i^\top B + b \right) A + a \right)$ and $\sigma(\bx_i)=\exp{(\bx_i)}/(\sum_{j=1}^N\exp{(\bx_i)})$ is the softmax function, a generalization of the logistic link function. The parameters $\st = (A, B, a, b)$ will be estimated using SGMCMC, where $A$, $B$, $a$ and $b$ are matrices of dimension: $100 \times 10$, $784 \times 100$, $1 \times 10$ and $1 \times 100$, respectively. We set normal priors for each element of these parameters
\begin{align*}
    A_{kl} | \lambda_A \sim N(0, 1), \quad B_{jk} | \lambda_B \sim N(0, 1), \quad a_l | \lambda_a \sim N(0, 1), \quad b_k | \lambda_b \sim N(0, 1),
\end{align*}
$    j = 1,\dots,784; \quad k = 1,\dots,100; \quad l = 1,\dots,10;$.

\subsubsection{Grid search}

We use the same grid as for PMF: \(\{-2. , -2.5, -3. , -3.5, -4. , -4.5, -5. , -5.5, -6. , -6.5, -7. , -7.5\}\). For SGHMC we also try two values of leapfrog steps: 5 and 10. 

We start from the MAP with Gaussian noise (scale: $\sigma=1$) and run $1,000$ iteration per grid point.

\subsubsection{MAMBA}

We use a time budget of $R=10sec$ (time of longest running sampler), and the same step size grid as for gridsearch: \(\{-2. , -2.5, -3. , -3.5, -4. , -4.5, -5. , -5.5, -6. , -6.5, -7. , -7.5\}\). We also use a grid for batch sizes: \(100\%, 10\%, 1\%\). For SGHMC we try two values of leapfrog steps: 5 and 10.

\subsubsection{Results}

We show in Table \ref{tab:NN_hyperparameters} the hyperparameters for the runs in Table \ref{tab:NN_uncertainty_tests}.

\begin{table*}
\caption{NN: Hyperparameters for the results in Table \ref{tab:NN_uncertainty_tests}. The batch size is given by $\tau$: the percentage of the total number of data. Namely: batch size $n = \lfloor{\tau N / 100}\rfloor$}
\label{tab:NN_hyperparameters}
\centering
\begin{tabular}{ p{2.5cm} p{1.5cm} p{2cm} p{2cm} p{2cm}}
  & & \textbf{MAMBA-KSD} &  \textbf{Grid Search} & \textbf{Heuristic}\\
\hline \\
 \textbf{SGLD} & $\log_{10}(h)$  & -5.5 & -3.5 &  -4.8\\
                 & $\tau$ ($\%$)  & 1 & 10 & 10 \\
 \midrule
 
 \textbf{SGLD-CV} & $\log_{10}(h)$  & -5.5 & -3.5 &  -4.8\\
     & $\tau$ ($\%$) & 1 & 10 &  10\\
                
 \midrule

  \textbf{SGHMC} & $\log_{10}(h)$  & -6.5 & -5 & -4.8 \\
     & $\tau$ ($\%$) & 1 &  10 &  10\\
     & $L$  & 5 & 10 & 10 \\
  \midrule
  \textbf{SGHMC-CV} & $\log_{10}(h)$  & -6 & -5 & -4.8 \\
     & $\tau$ ($\%$) &  1  & 10 &  10\\
     & $L$  & 5 & 10 & 10 \\
  
 \midrule
  \textbf{SGNHT}  & $\log_{10}(h)$  & -5 & -5 & -4.8 \\
     & $\tau$ ($\%$) & 1  & 10 &  10\\
  
   \midrule
  \textbf{SGNHT-CV}  & $\log_{10}(h)$  & -7.5 & -4.5 &  -4.8 \\
     & $\tau$ ($\%$) & 1  & 10 &  10\\
\end{tabular}
\end{table*}

\begin{table*}
\caption{Comparison of tuning methods for the neural network model. For each tuning method and each sampler we report the ECE and MCE (as percentages), as well as the test accuracy and the number of samples.}
\label{tab:NN_uncertainty_tests_full}
\centering
\begin{tabular}{ p{1.8cm} p{1.9cm} p{2.1cm} p{2.1cm} p{2.1cm} }
  & & \textbf{MAMBA-KSD} & \textbf{Grid Search} & \textbf{Heuristic} \\
\hline \\
 
 \textbf{SGLD} & ECE ($\%$) & 1.04 & 14.6 & \textbf{0.8} \\
 & MCE ($\%$) & 36.4 & 42.1 & \textbf{23.3} \\
 & Test acc. &  93.1 & \textbf{93.8} & 93.3  \\
 & \# of samples & 96,922 & 16,343 & 15,192\\
 \midrule
 
 \textbf{SGLD-} & ECE ($\%$) &  0.9 & 8.8 & \textbf{0.7}\\
 \textbf{CV} & MCE ($\%$) & \textbf{15.7} & 40.7 & 22.0 \\
 & Test acc. & 93.1 & \textbf{94.2} & 93.2 \\
 & \# of samples & 67,395 & 9,659 & 9,534 \\
 \midrule

  \textbf{SGHMC} & ECE ($\%$) & \textbf{0.7} & 20.1 & 50.9  \\
 & MCE ($\%$) & \textbf{47.1} & 65.5 & 71.6 \\
 & Test acc. & \textbf{93.0} & 92.5 & 91.7 \\
 & \# of samples & 23,671 & 1,761 & 1,717 \\
 \midrule
 
 \textbf{SGHMC-} & ECE ($\%$) & \textbf{0.7} & 25.1 & 40.8 \\
 \textbf{CV} & MCE ($\%$) & \textbf{21.3} & 55.2 & 74.8  \\
 & Test acc. & \textbf{93.1} & 82.9 & 90.1 \\
 & \# of samples & 15,327 & 1013 & 984 \\
 \midrule
 
 \textbf{SGNHT} & ECE ($\%$) & 9.3 & \textbf{5.4} & 6.2 \\
 & MCE ($\%$) &  45.7 & \textbf{42.2} & 43.2 \\
 & Test acc. & 94.0 & 95.1 &  \textbf{95.2} \\
 & \# of samples & 88,021 & 17,062 & 16,727 \\
 \midrule
 
 \textbf{SGNHT-} & ECE ($\%$) & \textbf{0.9} & 7.7 & 7.0 \\
 \textbf{CV} & MCE ($\%$) &  \textbf{27.4} & 42.3 & 51.5 \\
 & Test acc. & 93.1 & 94.6 & \textbf{95.0} \\
 & \# of samples & 62,389 & 9,372 & 9,382 \\
 
\end{tabular}
\end{table*}

\section{SAMPLERS}\label{sec:appendix_sgmcmc_samplers}

\subsection{SGHMC}\label{sec:appendix_sampler_detail_sghmc}

We use the recommended parameterization from \cite{chen2014stochastic} (see Equation 15 in that paper):

\begin{equation}
\begin{cases}
    \Delta \st &= v \\
    \Delta v &= -h \nabla \tilde{U}(\st) - \alpha v + \mathcal{N}(0, 2(\alpha - \hat{\beta}) h),
\end{cases}    
\end{equation}

with $v$ as the momentum variable, $\st$ the parameter of interest, $h$ the step size, and $\alpha$ the friction coefficient. The friction coefficient $\alpha$ and the noise estimation term $\hat{\beta}$ are tunable hyperparameters. We have set them to be $\alpha = 0.01$ and $\hat{\beta}=0$ for all the experiments.

\subsection{SGNHT}\label{sec:appendix_sampler_detail_sgnht}

From \cite{Ding2014_SHNHT}: we augment the parameter space with a momentum variable and a temperature variable. Here D is the dimension of the parameter. The tunable parameters are $h$ and $a$. We fix $a=0.01$) throughout.

\begin{equation}
\begin{cases}
  v_{n+1} &= v_n + h\hat{\nabla} \log \pi(x_n) - \alpha_n v_n + \sqrt{2a h}\xi \\
  x_{n+1} &= x_{n+1} + v_n \\
  \alpha_{n+1} &= \alpha_n + \frac{1}{D}v_{n+1}^Tv_{n+1} - h
  \end{cases}    
\end{equation}

\subsection{Control variates}

SGLD with control variates (SGLD-CV) uses the update in \eqref{eq:sgld}, but with an alternative estimate for the gradient $\hat{U}(\st)$ defined as follows. Let $\st_{MAP}$ denote the maximum a-posteriori (MAP) estimate of the posterior. The estimator for the gradient at parameter $\st$ is given by:

\begin{equation}\label{sgld_CV_gradient_estimate}
    \nabla \hat{U}(\st) = \nabla U(\st_{MAP}) + \left( \nabla \tilde{U}(\st)  - \nabla \tilde{U}(\st_{MAP}) \right)
\end{equation}

See \cite{Baker:2017} for more details on SGLD-CV. SGHMC-CV and SGNHT-CV are defined similarly.

\end{document}